\documentclass[aps,prb,twocolumn,superscriptaddress,showpacs]{revtex4}

\usepackage{graphicx}
\usepackage{dcolumn}
\usepackage{bm}

\begin{document}

\title{Ground state and magnetic phase transitions of orthoferrite DyFeO$_3$}

\author{Z. Y. Zhao}
\affiliation{Hefei National Laboratory for Physical Sciences at
Microscale, University of Science and Technology of China, Hefei,
Anhui 230026, People's Republic of China}

\author{X. Zhao}
\email{xiazhao@ustc.edu.cn}

\affiliation{School of Physical Sciences, University of Science
and Technology of China, Hefei, Anhui 230026, People's Republic of
China}

\author{H. D. Zhou}
\affiliation{Department of Physics and Astronomy, University of
Tennessee, Knoxville, Tennessee 37996-1200, USA}

\affiliation{National High Magnetic Field Laboratory, Florida
State University, Tallahassee, Florida 32306-4005, USA}

\author{F. B. Zhang}
\affiliation{Hefei National Laboratory for Physical Sciences at
Microscale, University of Science and Technology of China, Hefei,
Anhui 230026, People's Republic of China}

\author{Q. J. Li}
\affiliation{Hefei National Laboratory for Physical Sciences at
Microscale, University of Science and Technology of China, Hefei,
Anhui 230026, People's Republic of China}

\affiliation{School of Physics and Material Science, Anhui
University, Hefei, Anhui 230039, People's Republic of China}

\author{C. Fan}
\affiliation{Hefei National Laboratory for Physical Sciences at
Microscale, University of Science and Technology of China, Hefei,
Anhui 230026, People's Republic of China}

\author{X. F. Sun}
\email{xfsun@ustc.edu.cn}

\affiliation{Hefei National Laboratory for Physical Sciences at
Microscale, University of Science and Technology of China, Hefei,
Anhui 230026, People's Republic of China}

\author{X. G. Li}
\affiliation{Hefei National Laboratory for Physical Sciences at
Microscale, University of Science and Technology of China, Hefei,
Anhui 230026, People's Republic of China}

\affiliation{Department of Physics, University of Science and
Technology of China, Hefei, Anhui 230026, People's Republic of
China}

\date{\today}

\begin{abstract}

Low-temperature thermal conductivity ($\kappa$), as well as
magnetization ($M$) and electric polarization ($P$), of
multiferroic orthoferrite DyFeO$_3$ single crystals are studied
with $H \parallel c$. When the crystal is cooled in zero field,
$M$, $P$, and $\kappa$ all consistently exhibit irreversible
magnetic-field dependencies. In particular, with 500 mK $< T \le$
2 K, all these properties show two transitions at the first run of
increasing field but only the higher-field transition is present
in the subsequent field sweepings. Moreover, the ultra-low-$T$ ($T
<$ 500 mK) $\kappa(H)$ shows a different irreversibility and there is
only one transition when the field is swept both up and down. All
the results indicate a complex low-$T$ $H - T$ phase diagram
involving successive magnetic phase transitions of the Fe$^{3+}$
spins. In particular, the ground state, obtained with cooling to subKelvin temperatures in zero field, is found to be an unexplored phase.

\end{abstract}

\pacs{75.85.+t, 66.70.-f}

\maketitle

\section{Introduction}

Magnetic phase transition induced by magnetic field is an
outstanding phenomenon in the strongly-correlated electron systems
and is associated with many physical interests, such as the
unconventional superconductivity,\cite{Lake, Kang} the
non-Fermi-liquid behaviors,\cite{Pfau} and the
multiferroicity,\cite{Kimura} etc. Multiferroicity induced by spin
order has attracted much attention due to its large magnetoelectric (ME) coupling. The spin-current model or the inverse Dzyaloshinsky-Moriya (DM) interaction \cite{Spin_current} can explain well the
production of electric polarization ($P$) in the non-collinear
spin systems, such as the perovskite $R$MnO$_3$ ($R$ = rare
earth).\cite{Kimura, RMnO3-2,RMnO3-3} When the spins are aligned
collinearly, $P$ can also be formed through the exchange striction
mechanism, such as in $R$FeO$_3$,\cite{Tokunaga_DyFeO3,
Tokunaga_GdFeO3, SmFeO3} Ca$_3$CoMnO$_6$,\cite{Ca3CoMnO6} and
$R$Mn$_2$O$_5$.\cite{RMn2O5-1, RMn2O5-2, RMn2O5-3} In these
materials, the spin structures are playing a key role in the ME
coupling and the formation of the spontaneous electric
polarization. The rare-earth-based orthoferrites $R$FeO$_3$ have
received a lot of research interests in last several decades,
particularly in the manifestations of the spin structures and spin
re-orientations.\cite{Kimel-1, Kimel-2, Kimel-3, Tokunaga_RFeO3,
Yamaguchi}

In DyFeO$_3$, the Fe$^{3+}$ moments exhibit $G_xA_yF_z$ (Fe$\rm_I$) spin configuration in Bertaut's notation \cite{Bertaut} at room temperature,  that is, the main component of the magnetic moment is along the $a$
axis, accompanying with weak ferromagnetism (WFM) along the $c$
axis.\cite{Specific_heat, Magnetization} Upon lowering
temperature, the Fe$^{3+}$ spins undergo a Morin transition\cite{Morin}
at $T\rm{_M} = 50$ K. At this transition the spin configuration changes to $A_xG_yC_z$ (Fe$\rm_{II}$).\cite{Magnetization} Moreover, with an applied
magnetic field, $H > H\rm{_r^{Fe}}$, along the $c$ axis the
Fe$^{3+}$ spin configuration could change back to
Fe$\rm_I$.\cite{Tokunaga_DyFeO3, Mossbauer} With further lowering
temperature, the Dy$^{3+}$ spins develop a long-range antiferromagnetic (AF)
order below $T\rm{_N^{Dy}}$ = 4.2 K.\cite{Tokunaga_DyFeO3} In the
AF state, Dy$^{3+}$ spins are confined in the $ab$ plane and the
spin configuration can be expressed as $G_xA_y$ with the Ising
axis deviating about 33$^\circ$ from the $b$ axis.\cite{DyAlO3-1, DyAlO3-2} The spin-induced multiferroicity was observed only at $T < T\rm{_N^{Dy}}$ and when the spin flop of Fe$^{3+}$ moments is introduced by a $c$-axis
field.\cite{Tokunaga_DyFeO3} Due to the interaction between
Dy$^{3+}$ and Fe$^{3+}$ spins, the Dy$^{3+}$ spins shift towards
the layers of Fe$^{3+}$ with opposite spin directions and far away from
those with the same spin directions, resulting in a collective displacement
of the Dy$^{3+}$ ions and producing a spontaneous $P$. Therefore, both the
Dy$^{3+}$ and Fe$^{3+}$ spin structures are crucial for the ME
phenomenon. However, the spin structure of Fe$^{3+}$ at low
temperatures actually has not been determined. An obvious
inconsistency in the early studies is that the Mossbauer
spectroscopy suggested a $G_xG_y$ structure,\cite{GxGy} while the
electric polarization results suggested a $A_xG_yC_z$
structure.\cite{Tokunaga_DyFeO3} Moreover, the ground state and the low-$T$ magnetic phase transitions are actually not known since all the previous works had not studied the physics of DyFeO$_3$ at temperatures lower than 1.5 K.

Heat transport has been proved to be a useful probe for the
low-$T$ magnetic transitions.\cite{Sologubenko1, Sologubenko2,
Sun_DTN, Wang_HMO, Zhao_GFO, Zhao_NCO, Fan_DTO, Zhang_GdErTO} In
this work, we study the low-$T$ thermal conductivity ($\kappa$),
as well as the magnetization ($M$) and electric polarization, of
DyFeO$_3$ single crystals with $H \parallel c$ after a zero field
cooling (ZFC) process. In particular, the $P$ measurements can be done at temperatures below 1 K, while the $\kappa$ is measured at temperatures down to several tens of milli-Kelvin.  It is found that $M(H)$, $P(H)$, and
$\kappa(H)$ all consistently exhibit peculiar low-field
irreversible behaviors at $T >$ 500 mK with the irreversibility field
smaller than $H\rm{_r^{Fe}}$. At $T <$ 500 mK, $\kappa(H)$ shows a
different irreversibility with larger $\kappa$ in the field-up
process, which is in contrast with the case above 500 mK. The
results suggest a complex low-$T$ $H - T$ phase diagram involving
successive field-induced magnetic phase transitions of Fe$^{3+}$
spins.

\section{Experiments}

High-quality DyFeO$_3$ single crystals were grown by a
floating-zone technique in flowing oxygen-argon mixture with the
ratio of 1:4. The crystals were cut precisely along the
crystallographic axes after orientation by using back-reflection
x-ray Laue photographs. The sample for magnetization measurement
is rod-like and the dimension is 1.70 $\times$ 0.75 $\times$ 0.55
mm$^3$ with the length, width and thickness along the $b$, $c$,
and $a$ axis, respectively. The sample for electric polarization
measurement is plate-like and the wide face is perpendicular to
the $c$ axis with dimension of 2.06 $\times$ 2.14 $\times$ 0.13
mm$^3$. The dimension of the sample for thermal conductivity
measurement is 3.86 $\times$ 0.63 $\times$ 0.14 mm$^3$ and the
longest dimension is parallel to the $c$ axis.

Magnetization was measured by a SQUID-VSM (Quantum Design).
Electric polarization was obtained by integrating the displacement
current measured by an electrometer (model 6517B, Keithley) in a
$^3$He refrigerator and a 14 T magnet. $P(T)$ was measured at a
rate of about 2 K/min from 300 mK to 5 K. In order to stabilize
temperature in the $P(H)$ measurements, the sweeping-field rate
must be slower for lower temperature, which is 0.25, 0.2, 0.15,
and 0.1 T/min at 2, 1.4, 1, and 0.7 K, respectively. Thermal
conductivity was measured by using a ``one heater, two
thermometers'' technique and three different
cryostats:\cite{Sun_YBCO, Sun_DTN, Wang_HMO, Zhao_GFO, Zhao_NCO,
Fan_DTO, Zhang_GdErTO} (i) in a $^3$He-$^4$He dilution
refrigerator at temperature regime of 70 mK--1 K; (ii) in a $^3$He
refrigerator at 0.3--8 K, and (iii) in a pulse-tube refrigerator
for zero-field data at $T >$ 5 K. In all these measurements, the
magnetic fields were applied along the $c$ axis.

\section{Results and Discussion}

\subsection{Magnetization}

\begin{figure}
\includegraphics[clip,width=8.5cm]{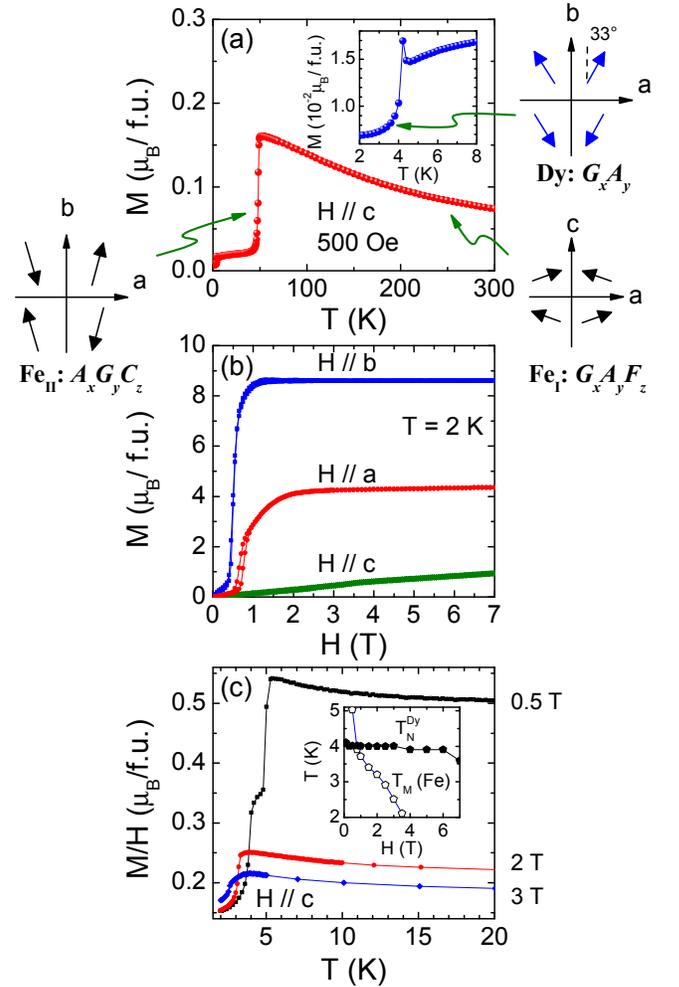}
\caption{(Color online) Magnetic properties of DyFeO$_3$ single
crystal. (a) Magnetic susceptibility in $H$ = 500 Oe along the $c$
axis. Inset: Zoom in of the low-$T$ data. The transitions at about 50 and 4.2 K correspond to the Morin transition of Fe$^{3+}$ and the N\'eel transition of Dy$^{3+}$, respectively. The schematics
illustrate the spin structures of Fe$^{3+}$ (bottom, black) and
Dy$^{3+}$ (top, blue). (b) Anisotropic magnetization at $T$ = 2 K.
(c) Representative data of the low-$T$ magnetic susceptibility in
different magnetic fields along the $c$ axis. Inset: field dependencies of the transition temperatures of the Dy$^{3+}$ N\'eel order and the Fe$^{3+}$ spin re-orientation, determined from the $M(T)$ curves.}
\end{figure}

The basic magnetic properties of DyFeO$_3$ are characterized by
the $M(T)$ and $M(H)$ measurements, of which the representative
data are shown in Fig. 1. These results are in good consistency
with the earlier works.\cite{Magnetization, Tokunaga_DyFeO3} The
temperature dependence of $M$ along the $c$ axis measured in $H$ =
500 Oe, shown in Fig. 1(a), has two transitions at 50 and 4.2 K.
The AF order of Fe$^{3+}$ spins is known to be formed at a high
temperature of $T\rm{_N^{Fe}} \sim$ 645 K, with a $G_xA_yF_z$
(Fe$\rm_I$) spin configuration at room temperature. The transition
at 50 K ($T\rm{_M}$) is a Morin transition, where the Fe$^{3+}$
structure changes to $A_xG_yC_z$
(Fe$\rm_{II}$).\cite{Magnetization} Another transition at 4.2 K
($T\rm{_N^{Dy}}$) corresponds to the AF ordering of Dy$^{3+}$
moments in the $G_xA_y$ configuration. The low-$T$ $M(H)$ curves
shown in Fig. 1(b) are consistent with these spin structures.

When the field is applied along the $c$ axis, the transition from
Fe$\rm_I$ to Fe$\rm_{II}$ shifts to lower temperature rapidly with
increasing field, but the AF order of Dy$^{3+}$ is robust against
the field, as seen in Fig. 1(c). It is known that Dy$^{3+}$
moments have strong anisotropy and are confined in the $ab$ plane;
therefore, the $c$-axis field can hardly to change either the
N\'eel transition or the Dy$^{3+}$ spin orientation. Thus, the
irreversible $M(H)$, $P(H)$, and $\kappa(H)$ behaviors at low
temperatures shown in the following sections are unambiguously
related to the successive magnetic transitions of Fe$^{3+}$ spins.

\begin{figure}
\includegraphics[clip,width=8cm]{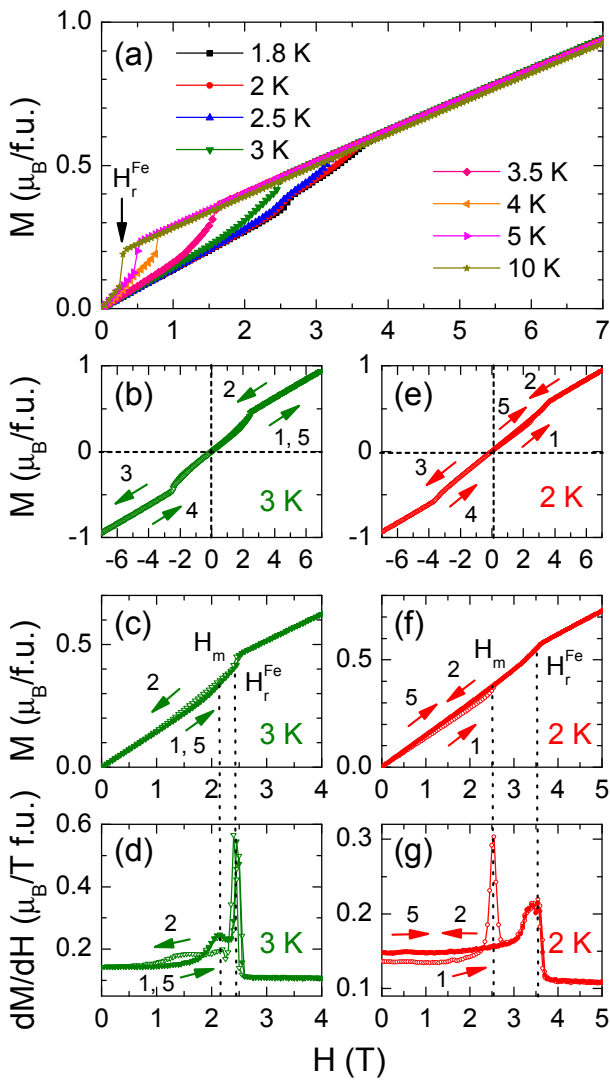}
\caption{(Color online) Magnetization of DyFeO$_3$ single crystal
with $H \parallel c$ after ZFC. Panel (a) shows the magnetization
in the positive field-up process. Other panels show full scan of
magnetization between 7 and -7 T, the magnetization curves in the
positive fields, and the differential curves with $T$ = 3 K (b-d)
and 2 K (e-g). The numbers and arrows denote the sweeping-field
sequences and directions. The open and solid symbols denote the
magnetization in the first full scan and the field-up process for
the second time. $H\rm{_r^{Fe}}$ and $H\rm_{m}$ represent two
transition fields, indicated by the peaks of the differential
curves.}
\end{figure}

Figure 2 shows the magnetization of DyFeO$_3$ single crystal with
$H \parallel c$ after ZFC. With $T > T \rm{_N^{Dy}}$, $M$ exhibits
an abrupt increase at $H \rm{_r^{Fe}}$, which corresponds to the
spin-flop transition of Fe$^{3+}$ moments from Fe$\rm_{II}$ to
Fe$\rm_I$.\cite{Tokunaga_DyFeO3, Mossbauer} With decreasing
temperature, this transition gradually moves to higher field but
becomes weaker, and finally evolves into a change of the slope
(Fig. 2(a)). With $T < T\rm{_N^{Dy}}$, there is a lower-field
transition shown by a peak at $H\rm_{m}$ in the differential
d$M$/d$H$ curves, besides the peak at $H \rm{_r^{Fe}}$, as shown
in Figs. 2(d) and 2(g). Moreover, a peculiar irreversible $M(H)$
behavior is observed at $H < H\rm{_r^{Fe}}$, with the
irreversibility field slightly larger than $H\rm_{m}$, as shown in
Figs. 2(b-g). Note that this irreversibility is repeatable at $T
>$ 2 K (Figs. 2(c) and 2(d)), which means that the field-up $M$ for the
second time (solid symbol, label 5) is identical to that for the
first field-up process (open symbol, label 1). However, with $T
\leq$ 2 K, the irreversibility becomes unrepeatable (Figs. 2(f)
and 2(g)); that is, the field-up $M$ for the second time is equal
to that in the field-down process (open symbol, label 2).
Therefore, there are two transitions in the field-up process and
only the higher-field one exists when the field is swept down.

\subsection{Electric Polarization}

\begin{figure}
\includegraphics[clip,width=8.5cm]{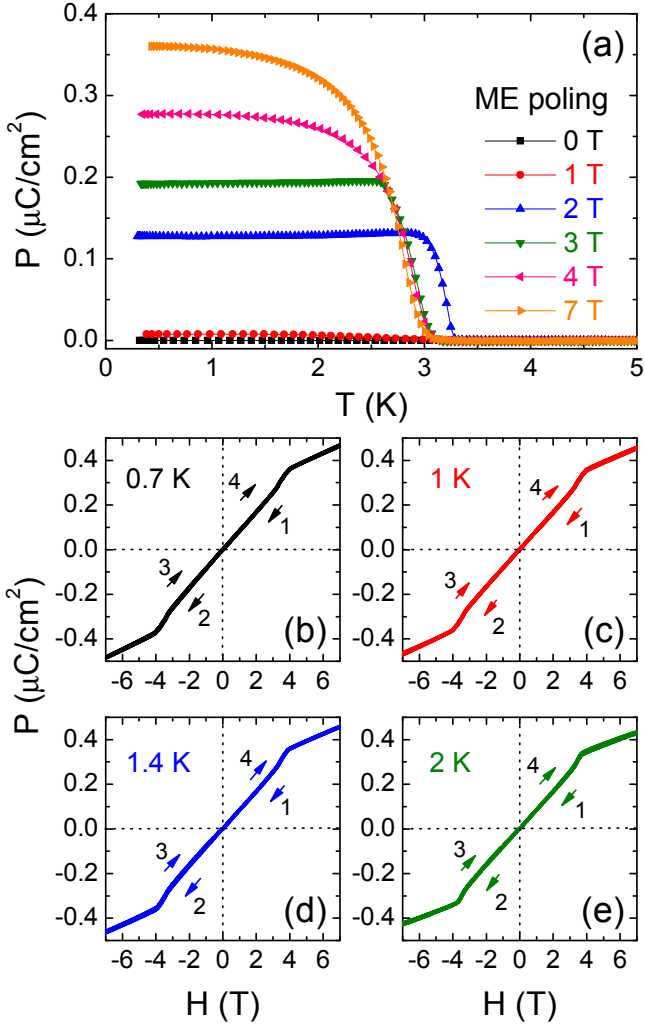}
\caption{(Color online) (a) Temperature dependencies of the
$c$-axis electric polarization of DyFeO$_3$ single crystal
measured after ME poling in magnetic field and $E = 3$ kV/cm
electric field, both along the $c$ axis. (b-e) Magnetic field
dependencies of $P$ measured after ME poling in 7 T and $E = 3$
kV/cm. The numbers and arrows denote the sweeping-field sequences
and directions.}
\end{figure}

The electric polarization along the $c$ axis of DyFeO$_3$ single
crystal is measured down to subKelvin temperatures. Figure 3(a) shows the temperature dependencies of $P$ measured after ME cooling from 5 K to 0.3 K with magnetic field and $E = 3$ kV/cm along the $c$ axis. $P$ is
temperature independent at very low temperatures and decreases to
zero at around $T\rm_N^{Dy}$, which indicates that the Dy$^{3+}$
spin order contributes to this electric polarization. The
magnitude of $P$ is enhanced and the transition temperature is
slightly suppressed with increasing field. Figures 3(b-e) shows
the magnetic-field dependence of $P$, measured after ME poling from 5 K in
7 T magnetic field and $E = 3$ kV/cm along the $c$ axis. At low
fields, $P(H)$ shows a linear ME behavior. With increasing field,
the spin-flop transition of Fe$^{3+}$ from Fe$\rm_{II}$ to
Fe$\rm_I$ causes a change of the slope of $P(H)$ at $H =
H\rm{_r^{Fe}}$, and the multiferroicity shows up. These results
are also in good agreement with the earlier
work.\cite{Tokunaga_DyFeO3}

\begin{figure}
\includegraphics[clip,width=8.5cm]{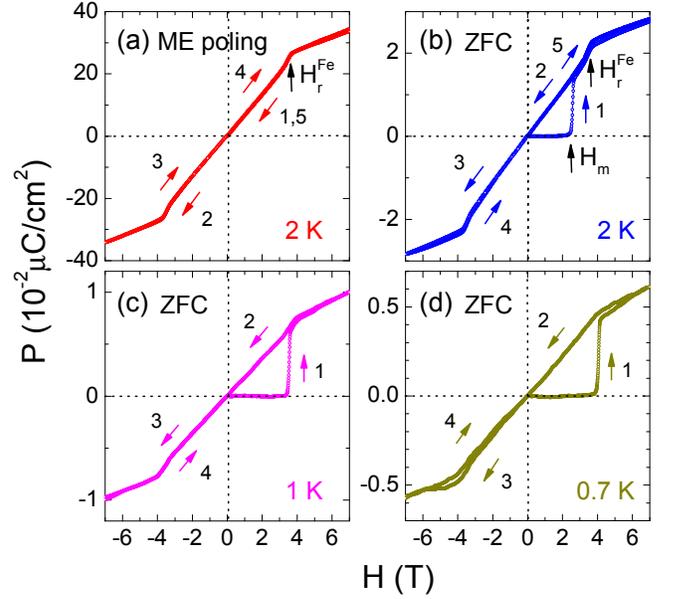}
\caption{(Color online) Magnetic-field dependencies of the
electric polarization of DyFeO$_3$ single crystal measured at (a)
2 K after ME poling and (b-d) at 2 -- 0.7 K after ZFC in $H
\parallel c$. The numbers and arrows denote the sweeping-field
sequences and directions. In panels (a) and (b), $H\rm_{m}$ and
$H\rm{_r^{Fe}}$ indicates two transition fields and are the same
as those from the magnetization.}
\end{figure}

However, $P(H)$ obtained with the ZFC process (cooling in zero
magnetic field and electric field) shows very different behavior
(Figs. 4(b-d)). First, the magnitude of $P$ after ZFC is one or
two orders smaller than that after ME poling, which is mainly due
to the electric domains. Second and more importantly, an
irreversible $P(H)$ is observed in the linear ME phase. At $T$ = 2
K (Fig. 4(b)), for example, $P$ is zero in low fields and exhibits
a step-like increase at $H\rm_{m} \approx$ 2.5 T for the first
field-up process. In higher-field, $P(H)$ is reversible and linear
with field. Note that $H\rm_{m}$ increases with decreasing
temperature and approaches $H\rm{_r^{Fe}}$ at 0.7 K (Figs.
4(b-d)). When $H > H\rm{_r^{Fe}}$ ($\approx$ 3.5 T), the system
enters the multiferroic phase due to the spin-flop transition of
Fe$^{3+}$ sublattice.\cite{Tokunaga_DyFeO3} The values of
$H\rm_{m}$ and $H\rm{_r^{Fe}}$ observed from $P(H)$ are
essentially consistent with those of $M(H)$. This consistency
suggests a common origin for the observed irreversibility of
$M(H)$ and $P(H)$.

\subsection{Thermal Conductivity}

\begin{figure}
\includegraphics[clip,width=6.5cm]{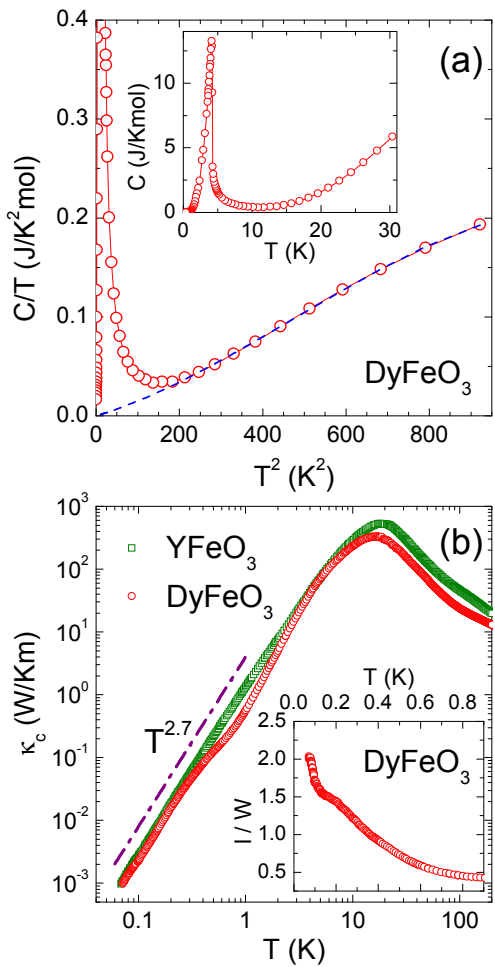}
\caption{(Color online) (a) The specific heat of DyFeO$_3$ single
crystal below 30 K, plotted in $C/T$ vs $T^2$. The inset displays
all the data at $T \le$ 30 K, which show a sharp peak at 4.1 K.
The dashed line shows the fitting to the high-$T$ data by using
the formula of phonon specific heat, that is, $C = \beta T^3 +
\beta_5 T^5 + \beta_7 T^7$. (b) Zero-field thermal conductivity of
DyFeO$_3$ with heat current along the $c$ axis. For comparison,
the data of YFeO$_3$ is also presented. The sample sizes are 3.86
$\times$ 0.63 $\times$ 0.14 mm$^3$ and 4.40 $\times$ 0.66 $\times$
0.16 mm$^3$ for DyFeO$_3$ and YFeO$_3$, respectively. The dash-dot
line indicates a $T^{2.7}$ temperature dependence. The inset shows
the temperature dependence of the phonon mean free path $l$
divided by the averaged sample width $W$.}
\end{figure}

The thermal conductivity of DyFeO$_3$ single crystal is studied
for probing the magnetic transitions at even lower temperatures
that our $M(H)$ and $P(H)$ measurements cannot be done.

Before presenting the heat transport results of DyFeO$_3$ crystal,
we show in Fig. 5(a) the low-$T$ specific heat data. A large and
sharp peak is observed at 4.1 K, which is related to the N\'eel
transition of Dy$^{3+}$ spins. It can be seen that the magnetic
contributions to the specific heat are important only at very low
temperatures and are likely to be negligible above $\sim$ 12 K,
where the specific heat data show a minimum. Therefore, the phonon
specific heat can be estimated from the high-$T$ data in Fig.
5(a). It is known that in the temperature range 0.02 $< T /
\theta_D <$ 0.1 ($\theta_D$ is the Debye temperature), the phonon
specific heat follows a low-frequency expansion of the Debye
function, $C = \beta T^3 + \beta_5 T^5+\beta_7 T^7 + ...$, where
$\beta$, $\beta_5$ and $\beta_7$ are temperature-independent
coefficients.\cite{Tari} It is found that this formula fits well
to the experimental data at $T >$ 15 K, as shown in Fig. 5(a), with
the fitting parameters $\beta = 1.28 \times 10^{-4}$ J/K$^4$mol,
$\beta_5 = 2.49 \times 10^{-7}$ J/K$^6$mol and $\beta_7 = -1.75
\times 10^{-10}$ J/K$^8$mol.

Figure 5(b) shows the temperature dependencies of the thermal
conductivity down to 70 mK in zero field, with the heat current
$J\rm_H$ applied along the $c$ axis. For comparison, the data for
YFeO$_3$ single crystal are also taken in the same temperature
regime. Note that the Y$^{3+}$ ions are nonmagnetic and there is
only AF order of Fe$^{3+}$ ions. YFeO$_3$ actually shows a simple
and pure phonon heat transport phenomenon at low temperatures: (i)
the $\kappa(T)$ curve exhibits a very large peak at about 20 K,
with the peak value of 520 W/Km, indicating a very high quality of
the single crystal; (ii) the temperature dependence of $\kappa$ is
roughly $T^{2.7}$ at subKelvin temperatures, which is close to the
$T^3$ boundary scattering limit of phonons.\cite{Berman} The
DyFeO$_3$ data are rather comparable to those of YFeO$_3$, except
for two notable differences. At first, the $\kappa$ of DyFeO$_3$
is smaller at high temperatures, although a phonon peak of 330
W/Km is also exceptionally larger for transition-metal oxides.
More remarkably, there is a clear concave structure in the
$\kappa(T)$ curve in the temperature regime of 0.3--3 K. A similar
result has been found in another orthoferrite
GdFeO$_3$.\cite{Zhao_GFO} It is clear that at $T < T\rm_N^{Dy}$,
the magnon excitations from the Dy$^{3+}$ spin system can have a
significant scattering on phonons, which results in a downward
deviation from the $T^{2.7}$ behavior. However, with lowering
temperature further, the $\kappa$ recovers to the $T^{2.7}$
dependence at $T <$ 300 mK. This means that the magnon scattering
effect is gradually smeared out. Apparently, the magnon
spectra has a finite energy gap (for example, originated from the
spin anisotropy), which prevents the low-energy magnons from being
thermally excited at very low temperatures.

It is possible to estimate the mean free path of phonons at low
temperatures and to judge whether the phonons are free from
microscopic scattering at subKelvin temperatures. The phononic
thermal conductivity can be expressed by the kinetic formula
$\kappa_{ph} = \frac{1}{3}Cv_pl$,\cite{Berman} where $C = \beta
T^3$ is the phonon specific heat at low temperatures, $v_p$ is the
average velocity and $l$ is the mean free path of phonons. Using
the $\beta$ value obtained from the above specific-heat data, the
phonon velocity can be calculated and then the mean free path is
obtained from the $\kappa$.\cite{Zhao_GFO, Zhao_NCO} The inset to
Fig. 5(b) shows the ratio between $l$ and the averaged sample
width $W = 2\sqrt{A/\pi}$ = 0.335 mm,\cite{Zhao_GFO, Zhao_NCO,
Berman} where $A$ is the area of cross section. It can be seen
that $l / W$ increases with lowering temperature and becomes larger than
one at lowest temperatures, which indicates that all the
microscopic phonon scatterings (including magnon scattering) are
negligible and the boundary scattering limit is established.\cite{Berman}

\begin{figure*}
\includegraphics[clip,width=18cm]{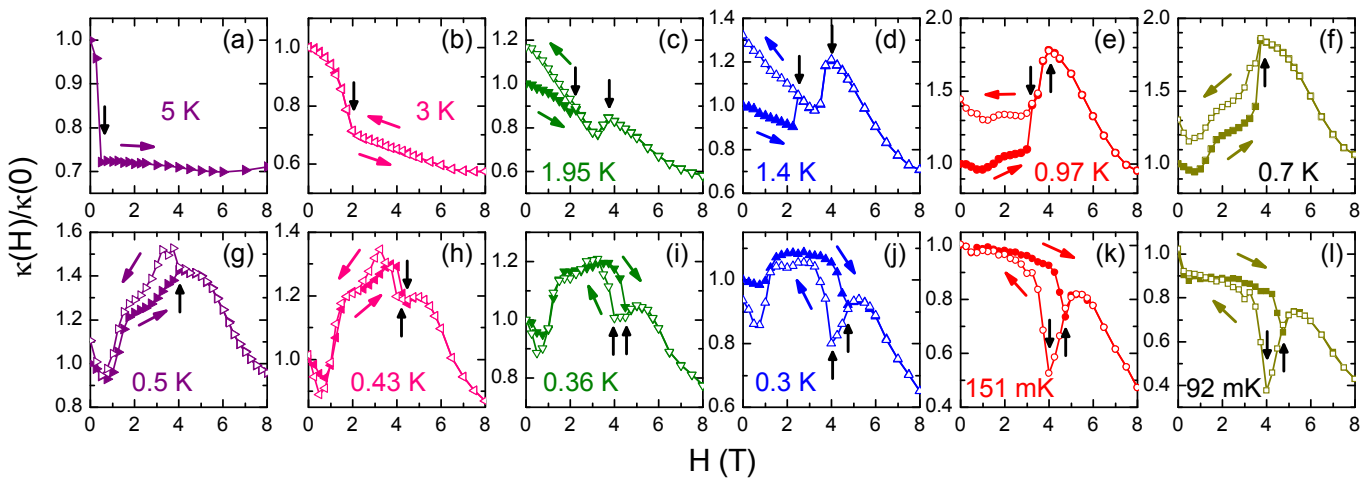}
\caption{(Color online) Magnetic-field dependencies of the
$c$-axis thermal conductivity of DyFeO$_3$ single crystal in $H
\parallel c$ after ZFC. As indicated by the colored arrows, the
data shown with solid symbols are measured in the field-up
process, while the open symbols show the data in the field-down
process. The black upright arrows indicate the characteristic
fields in $\kappa(H)$ isotherms.}
\end{figure*}

The field dependence of thermal conductivity after ZFC is measured
down to 92 mK and the results are presented in Fig. 6. With
decreasing temperature, $\kappa$ exhibits complicated field
dependence: (i) with $T>$ 2 K, $\kappa(H)$ is reversible and shows
a sharp decrease at $H \rm{_r^{Fe}}$; (ii) with 0.7 K $\leq T
\leq$ 2 K, $\kappa(H)$ shows an analogous irreversible behaviors
to those of $M(H)$ and $P(H)$. There are two maximums in the
field-up process and only the high-field one is kept when
decreasing field, resulting in a larger $\kappa$ in the field-down
process at low fields; (iii) when temperature is further decreased
down to below 500 mK, $\kappa(H)$ exhibits a different
irreversible behavior; that is, there is only one dip in the
sweeping-field run with a higher location in the field-up process.
One notable feature is that the $\kappa$ in the field-up process
becomes larger than that in the field-down process, which is
opposite to the case above 500 mK.

It has been known from many previous studies on the
antiferromagnetic materials that the thermal conductivity could
exhibit drastic changes at the magnetic phase transitions or the
spin re-orientations.\cite{Sologubenko1, Sologubenko2, Sun_DTN,
Wang_HMO, Zhao_GFO, Zhao_NCO, Fan_DTO, Zhang_GdErTO} The
mechanisms are mostly due to the sudden changes in the population
of the magnetic excitations, which can effectively scatter
phonons.\cite{Wang_HMO, Zhao_GFO, Zhao_NCO, Fan_DTO, Zhang_GdErTO}
In some rare case, the magnons well populated at the spin-flop
transition can also contribute to the heat transport by acting as
heat carriers.\cite{Zhao_NCO} In present work, the anomalies of
the $\kappa(H)$ curves can be apparently attributed to the
transitions of magnetic structures of DyFeO$_3$, which is also a
common phenomenon to GdFeO$_3$,\cite{Zhao_GFO}. However, the
hysteresis of $\kappa(H)$ are rather different between these two
materials.

\subsection{Phase Diagram and Magnetic Transitions}

Considering the above experimental results, one can find a common
feature for the ZFC $M(H)$, $P(H)$, and $\kappa(H)$ data, that is,
an irreversible behavior is observed only in the first field-up
process, and it disappears in the following sweeping-field
process. One possible reason for the observed irreversibility is
the multi-domain effect in the multiferroic
materials.\cite{Tokunaga_GdFeO3} Since the crystal is cooled in
zero field without ME poling, at zero and low fields, the
macroscopic polarization might be zero due to the compensation
among different electric domains. With increasing field, the
sample gradually changes into a fewer-domain state and the finite
magnitude of $P$ appears, which could be accompanied with some
change in the $M(H)$ curve. However, from our results at $T \leq$ 2 K, the $P(H)$ and $M(H)$ curves are reversible in the subsequent sweeping-field
process except for the first one. This is
obviously different from the case of the $P(H)$/$M(H)$ loop in
usual ferroelectric/ferromagnetic materials, in which the loop
always exists due to the presence of domains. On the other hand,
if the domain effect is the major reason for the observed
irreversibility, the phonon scattering by the domain walls should
be effective in the whole low-$T$ range. In the case of
GdFeO$_3$,\cite{Tokunaga_GdFeO3} when the crystal is cooled in zero field, the field-up $P$ is larger than that in the field-down run,
which indicates that there are less ferroelectric domains with
increasing field. The larger $\kappa(H)$ in the field-up process
can be attributed to the weak phonon scattering by the domain
walls.\cite{Zhao_GFO} Similar to the case of GdFeO$_3$, the
$\kappa(H)$ at $T >$ 500 mK behaves consistently with the $M(H)$ and
$P(H)$ curves in DyFeO$_3$; that is, the ME domain walls are
likely playing a role in scattering phonons. However, upon further
cooling ($\le$ 300 mK) the $\kappa$ in the field-up process
becomes larger than that in the field-down process, which can not
be explained by the domain wall scattering on phonons. In a word,
although the multi-domain effect could have influence on the
magnitudes of $M(H)$, $P(H)$, and $\kappa(H)$, when the crystal is
cooled without ME poling, it seems to be not the dominant factor
for the observed irreversible behaviors.

The second possible reason for the step-like increase of $P$ and
the consistent low-field irreversibility shown by $M(H)$, $P(H)$,
and $\kappa(H)$ is the presence of an unknown zero-field magnetic
structure below $T\rm_N^{Dy}$ after ZFC, which is metastable in
the field and disappears when the field is swept down. Actually,
it is a rather common feature of multiferroic materials that a
metastable spin configuration is formed in ZFC process while another
spin structure is stabilized after ME cooling. A recent example is
the field-induced metastable phase in hexaferrites, observed by a
neutron scattering.\cite{Lee} This
scenario can easily explain the $P(H)$ and $M(H)$ behaviors. Due
to the strong out-of-plane anisotropy of Dy$^{3+}$ spins, the
applied field along the $c$ axis could not change their spin
arrangements directly, as the magnetization in Fig. 1(c)
indicated. Most likely, only the change of Fe$^{3+}$ spin
structure is involved. Therefore, the zero-field structure of
Fe$^{3+}$ at $T < T\rm{_N^{Dy}}$ should be different from the
known Fe$\rm_{II}$ at $T > T\rm{_N^{Dy}}$ and is hereafter named
as Fe$\rm_{III}$. It is also notable that the repeatable
irreversibility begins to appear with $T < T\rm{_N^{Dy}}$ and
becomes unrepeatable with $T \le$ 2 K, the transition from
Fe$\rm_{II}$ to Fe$\rm_{III}$ is therefore a gradual process
between $T\rm{_N^{Dy}}$ and 2 K. Another notable feature is
that the transition from Fe$\rm_{II}$ to Fe$\rm_I$ at $T >
T\rm{_N^{Dy}}$ could cause a sharp increase of $M$. Due to the
presence of irreversibility, this feature gradually vanishes at $T
< T\rm{_N^{Dy}}$; instead, two successive magnetic transitions are
present, and the higher-field one is from Fe$\rm_{II}$ to
Fe$\rm_I$, which is the same as the case at high temperatures.

Based on the above discussions and experimental data of $M(H)$,
$P(H)$, and $\kappa(H)$, a low temperature $H - T$ phase diagram
with $H \parallel c$ and in the case of ZFC is constructed, as
shown in Fig. 7. We propose that besides the known Fe$\rm_{I}$ and
Fe$\rm_{II}$, the phase diagram involves another magnetic phase,
Fe$\rm_{III}$. The most important feature is that the Fe$^{3+}$
spins undergo successive transitions: Fe$\rm_{III} \rightarrow$
Fe$\rm_{II} \rightarrow$ Fe$\rm_I$ with increasing field and at 500 mK
$< T < T\rm_N^{Dy}$, and only the transition from Fe$\rm_{I}
\rightarrow$ Fe$\rm_{II}$ is present when the field is decreased,
which is consistent with the FC results. It is known that there
are four Fe$^{3+}$ spin configurations that could produce linear
ME response, i.e., $G_xA_yF_z$ (Fe$\rm_I$), $A_xG_yC_z$
(Fe$\rm_{II}$), $F_xC_yG_z$, and
$C_xF_yA_z$.\cite{Magnetic_symmetry} Since the crystal is cooled
without ME poling, it is in principal difficult to judge from the
electric polarization whether there is spontaneous $P$ or not in
zero field, and accordingly to judge the definite magnetic structure of
Fe$\rm_{III}$. This is because that the sample after ZFC could be
in a multi-domain phase with spontaneous polarization, even though
the measured $P$ is zero. However, since the low-$T$ $P(H)$ loops
actually behave rather differently from the usual $P(H)$
hysteresis curves, some kind of magnetic-structure transition
other than the multi-domain effect may play a more important role. Nevertheless, we have to leave the definition of the Fe$\rm_{III}$ magnetic structure as an open question, and further investigations like neutron scattering studies on single crystals are called for.

\begin{figure}
\includegraphics[clip,width=7.5cm]{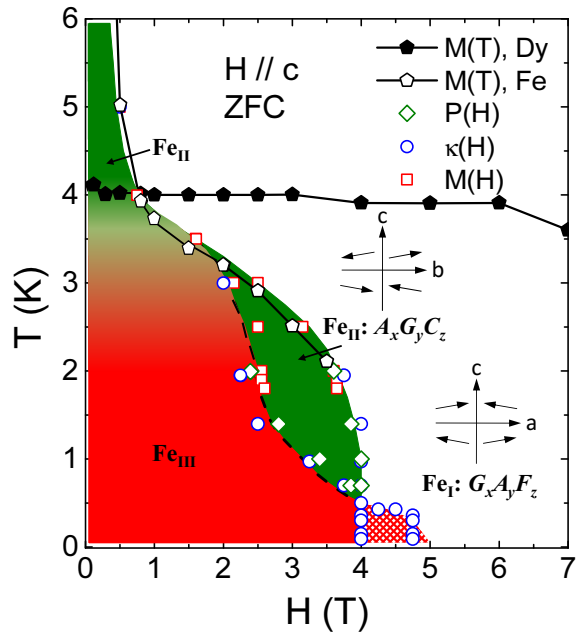}
\caption{(Color online) ZFC $H-T$ phase diagram of DyFeO$_3$
obtained from the $M(H)$, $P(H)$, and $\kappa(H)$ data. The
transition from Fe$\rm_{II}$ to Fe$\rm_{III}$ is actually a
gradual change, as a continuous change of color indicates. Above
500 mK, Fe$\rm_{III}$ is a metastable state; that is, it exits
when sweeping-up field to the boundary indicated by the dashed
line but does not appear when sweeping-down field. Below 500 mK,
Fe$\rm_{III}$ becomes stable in sweeping field. The crossed red
area represents an irreversible region due to the first-order
transition from Fe$\rm_{III}$ to Fe$\rm_I$. The transition
temperatures of the N\'eel order and the spin re-orientation for
Dy$^{3+}$ and Fe$^{3+}$, determined from the $M(T)$ curves (Fig.
1(c)), are also shown.}
\end{figure}

This phase diagram could explain the behavior of $\kappa(H)$ down
to the lowest temperature. (i) With $T > T\rm{_N^{Dy}}$, when the
field is applied, the Fe$^{3+}$ spin structure changes from
Fe$\rm_{II}$ to Fe$\rm_I$. Accordingly, a sharp decrease of
$\kappa$ is observed, due to the phonon scattering by the magnetic
excitations.\cite{Wang_HMO, Zhao_GFO, Fan_DTO} (ii) With $T \le$ 2
K, the applied field could induce successive transitions from
Fe$\rm_{III}$ to Fe$\rm_{II}$ and then to Fe$\rm_I$ in the
field-up process, and $\kappa(H)$ shows two local maximums due to
the weakened scattering of phonons across each transition. The
transition from Fe$\rm_{II}$ to Fe$\rm_{III}$ is absent when the
field is decreased, resulting in a larger $\kappa$ below
$H\rm_{m}$. (iii) With further cooling, $H\rm_{m}$ becomes larger
and there is only a first-order transition from Fe$\rm_{III}$ to
Fe$\rm_I$ directly with $T <$ 500 mK, which results in a different
field dependence of $\kappa$. It should be noted that below 500 mK
the Fe$\rm_{III}$ state is always present in the sweeping-field
process. This indicates that the Fe$\rm_{III}$ state, which is
metastable in sweeping magnetic field above 500 mK, becomes stable
at $T <$ 500 mK.

In passing, it should be pointed out that, besides the magnon
excitations, the paramagnetic spins could also have influence on
the phonon thermal conductivity.\cite{Berman,
paramagnetic_scattering-1, paramagnetic_scattering-2} Due to the
paramagnetic scattering effect, $\kappa(H)$ usually shows a
dip-like behavior. From Figs. 6(e-j), the dip around 1 T may be
ascribed to the paramagnetic scattering on phonons. The
complicated field dependencies at low fields between 0.97 K and
300 mK could be regarded as a superposition of magnon
scattering at $H\rm_m$ on the recovery of $\kappa$ due to the
paramagnetic scattering, as indicated by the shoulder-like feature
around 2 T for both field-up and field-down processes.

\section{Summary}

Our detailed results of magnetization, electric polarization, and
thermal conductivity point to a complex magnetic phase diagram of
DyFeO$_3$ involving successive field-induced magnetic phase
transitions of Fe$^{3+}$ spins. First, the unknown ground state at
ultra-low temperatures (Fe$\rm_{III}$) is likely to have no
linear ME effect. Apparently, this ground state is determined by
the interaction between Dy$^{3+}$ and Fe$^{3+}$ spins. In
particular, when the Dy$^{3+}$ sublattice orders
antiferromagnetically at $T\rm{_N^{Dy}}$, the Fe$^{3+}$ spins have
to change their orientation. Second, with 500 mK $< T \le$ 2 K, a
low-field irreversibility of the magnetic transitions is observed
when the sample is cooled in zero field, which indicates that
Fe$\rm_{III}$ structure is metastable with sweeping magnetic
field. A similar irreversible behavior is
probably the one found in another orthoferrite TbFeO$_3$, probed
by the capacitance measurement,\cite{TbFeO3} in which an unknown
low-field phase also disappeared with sweeping-down field. It
seems that only at very low temperatures ($<$ 500 mK), the
interaction between Dy$^{3+}$ and Fe$^{3+}$ spins is strong enough
to stabilize the Fe$\rm_{III}$ structure. The further studies of
these materials and probably other members of orthoferrite
$R$FeO$_3$ could help to explore their complicated low-$T$
magnetic structures and related interesting ME effect.

\begin{acknowledgements}

This work was supported by the National Natural Science Foundation
of China, the National Basic Research Program of China (Grants No.
2009CB929502, 2011CBA00111, and 2012CB922003), and the Fundamental
Research Funds for the Central Universities (Programs No.
WK2340000035 and WK2030220014).

\end{acknowledgements}

\end{document}